\documentclass{PoS}
\usepackage{epsfig}

\newcommand{\be}{\begin{equation}} 
\newcommand{\en}{\end{equation}}
\newcommand{\bea}{\begin{eqnarray}}
\newcommand{\ena}{\end{eqnarray}}

\newcommand{\hbo}{\hbox to 1 true cm {\hfill } } 
\newcommand{\tr}{\hbox{tr}}

\PoS{PoS(LAT2005)201}

\title{Zero density heavy quark SU(2) gauge theory and the Stefan-Boltzmann 
limit }

\ShortTitle{Zero density heavy quark SU(2) gauge theory }

\author{\speaker{Kurt Langfeld}, Norbert Lages, and Hugo Reinhardt \\ 
        Institute for Theoretical Physics, University of Tuebingen, 
        Germany \\
        E-mail: \email{kurt.langfeld@uni-tuebingen.de}}

\abstract{
SU(2) lattice gauge theory is investigated where the traces of the
Wilson lines at any lattice point and along each
direction is constrained to zero. Hence, each of the
lattice configurations possesses  a vanishing density of heavy
(anti-) quarks. The results are compared with those of pure
SU(2) gauge theory which can be interpreted as the grand canonical
realization of the heavy quark theory where only the ensemble
average of the heavy quark density vanishes.

The static quark anti-quark potential of the constrained theory
is obtained from (spatially smeared) Wilson loops at zero temperature.
We find that the potential coincides with that of pure SU(2)
gauge theory (without constraints).
Hence, the familiar "running" of the lattice spacing
with $\beta $ is recovered.
}

\FullConference{XXIIIrd International Symposium on Lattice Field Theory\\
                 25-30 July 2005\\
                 Trinity College, Dublin, Ireland}

\begin{document}

\section{The role of the Polyakov line}

\subsection{ Heavy quark physics } 

If the quark determinant is calculated for an arbitrary gluon 
field in the heavy quark limit, the correction to the pure gluonic 
action density is given by~\cite{Langfeld:1999ig}: 
$$ 
s_\mathrm{heavy}(x) \propto e^{ \frac{\mu - m }{T} } \; \tr P(x)
+ e^{ \frac{- \mu - m }{T} } \; \tr P^\dagger (x) \; ,
$$ 
where $m$ is the heavy quark mass, $T$ is the temperature, and $\mu $ 
is the quark chemical potential. $P(x)$ is the Polyakov line, i.e., 
the product of all time-like links forming a line parallel to 
Euclidean time. 

\vskip 0.3cm 
Taking the derivative of the partition function with respect to the 
chemical potential provides the expectation value of the baryon density. 
At vanishing chemical potential, we find 
$$ 
\langle \rho \rangle \; \propto \; \left\langle \Im \;  \tr P(x) 
\right\rangle \; . 
$$ 
In the case of SU(2), this density vanishes because the trace of 
any SU(2) matrix is real. For $SU(N>2)$, $\langle \rho \rangle $ 
vanishes if the center symmetry is realized: If the center transformation 
$$ 
P(x) \rightarrow P^\prime (x) \; = \; z \; P(x) \; , \hbo 
z = \exp \left\{ i \frac{2\pi }{N} m \right\} \; , \hbo 
-N < m \le N \; , 
$$
is realized, real part and imaginary parts of $ \langle  \tr P(x) 
\rangle $ vanish. While a particular lattice configuration 
makes a non-trivial contribution to $ \tr P(x) $, the ensemble 
average causes its expectation value to vanish (see figure~\ref{fig:2} 
for an illustration for the case SU(2)). This is 
no longer the case if $T$ exceeds the critical 
temperature $T_c$ for deconfinement: In the latter case, the 
above center symmetry is spontaneously broken. The 
ground state suggested by the lattice ensembles seems to be populated 
with heavy quarks of $N$-ality $m_s$ (where $m_s$ specifies the 
realized center sector). 

\vskip 0.3cm 
Lattice configurations of a standard simulation of pure $SU(N)$ gauge 
theory with periodic boundary conditions can be interpreted as 
part of a grandcanonical ensemble where, on average, the density 
of heavy baryons vanishes. By contrast, if we demand that 
$$ 
\Im \;  \tr P(x) \; = \; 0 
$$
for each configuration, we sample a canonical ensemble for which 
the baryon density is identical zero for each ensemble member. 

\subsection{ Topological sectors of perturbative vacua } 
\begin{figure}
\centerline{
\epsfig{ file = 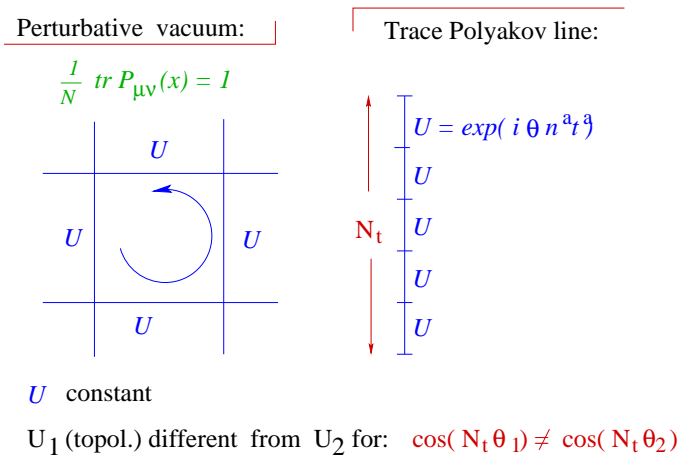, 
width = 0.6 \textwidth } 
}
\caption{ Topological different perturbative sectors on a torus 
for a SU(2) gauge theory. } 
\label{fig:1} 
\end{figure}
It was pointed out by M.~Schaden in~\cite{Schaden:2004ah} that 
even the Yang-Mills groundstate can fall into topological different 
classes on a torus. Let us briefly review his arguments. 

\vskip 0.3cm 
A perturbative ground state of SU(N) lattice Yang-Mills theory 
with Wilson action is given by a link configuration with 
\be 
\frac{1}{N} \; \tr P_{\mu \nu }(x) \; = \; 1 \; , \hbo 
\forall x, \, \nu > \mu \; , 
\label{eq:1} 
\en 
where $ P_{\mu \nu }(x)$ is the plaquette built from the links 
$U_\mu (x) $: 
$$ 
P_{\mu \nu }(x) \; = \; U_\mu (x) \; U_\nu (x+\mu) \; 
U^\dagger \mu (x+\nu) \; U^\dagger _\nu (x) \; . 
$$
Obviously, the condition (\ref{eq:1}) is satisfied for all 
constant links 
$$ 
U_\mu (x) \; = \; U \; =: \; \exp \{ i \theta ^a t^a \}, 
\hbo \forall x, \mu 
$$ 
and their gauge equivalents (see figure~\ref{fig:1}). 
Let us consider the Wilson line $P$ in time direction (Polyakov line) 
for a lattice 
with $N_t$ links in time direction. Since the trace of the 
Polyakov line is gauge invariant for a lattice gauge theory 
with periodic boundary conditions, the perturbative vacua, specified 
by $U$, can be classified by 
\be 
\kappa (U) \; = \; \frac{1}{N} \tr P \; = \; \frac{1}{N} \tr 
\exp \{ i N_t \theta ^a t^a \} \; . 
\label{eq:2} 
\en 
In particular, two candidates $U_1$ and $U_2$ are topological 
inequivalent if $\kappa (U_1) \not= \kappa (U_2)$.

\section{ Constrained Yang-Mills theory } 

\subsection{ Theory } 

In the following, we will show results from a simulation of  $SU(2)$ and 
$SU(3)$ lattice gauge theory on a 4-torus using lattice configurations with 
periodic boundary conditions. The lattice configurations 
are subjected to the constraints: 
\bea 
\tr P(\vec{x}) =0, \; \; \; & \forall &  \vec{x} \; \; \; \; \; 
\hbo \mathrm{for} \; \; \; SU(2) 
\\ 
\Im \, \tr P(\vec{x}) =0, \; \; \; & \forall &  \vec{x} \; \; \; \; \; 
\hbo \mathrm{for} \; \; \; SU(3) \; . 
\label{eq:csu3}
\ena 
Results will be shown for temperatures below and above the 
deconfinement temperature. 

\vskip 0.3cm
The results are interesting for the following reasons: (i) 
The results will show whether YM-theory with heavy quark 
fluctuations suppressed (i.e.~constrained YM-theory) yields the 
same results as pure Yang-Mills theory with periodic boundary 
conditions. Due to spontaneous center symmetry breaking at high 
temperatures, differences are expected at least for the deconfinement 
phase. (ii) If constrained  and standard YM-theory yield the 
same physics at least in the confinement phase, the range 
of applicability of Schaden's center symmetric perturbation 
theory~\cite{Schaden:2004ah} might be even extended to SU(2) 
gauge theory. (iii) If constrained YM-theory is confining, 
the confinement mechanism might well be in the reach of a semi-classical 
treatment of the partition function based upon background fields with 
particular holonomies such as calorons~\cite{Kraan:1998sn,Bruckmann:2004nu}. 

\vskip 0.3cm
To be specific, the partition function of the simulation is e.g.~in 
the SU(2) case given by 
\be 
Z \; = \; \int {\cal D}U \; \; \prod _{\vec{x}} \delta \Bigl(
\tr P(\vec{x}) \Bigr) \; \; \; 
\exp \Bigl\{ S_\mathrm{Wil} \Bigr\} \; . 
\label{eq:z}
\en 
One might ask whether the constraints in the 
$\delta $-functions affect the bulk properties of the lattice 
ensemble at all: $N_t$ link elements contribute to a 
given Polyakov line. Hence, $3 \, N_t $ real numbers (for SU(2)) 
must satisfy one constraint. This seems to suggest that 
the constraint is irrelevant in the infinite volume limit. 
Note, however, that $N_t -1 $ links can be gauged to unity. 
In addition, the remaining link can be chosen diagonal (up to a gauge 
transformation). This implies that in lattice gauge theory, there is 
only one physical degree of freedom contained in the $N_t$ links 
of the Polyakov line. From this arguments, we expect a significant 
influence from the constraints in (\ref{eq:z}). 

\subsection{ Numerical setup } 

Firstly, a local update algorithm on the basis of M.~Creutz 
heat bath algorithm was implemented. There, a particular 
link was chosen for the update, and the update of this link 
was performed according to its probabilistic weight, properly taking 
into account the constraint. 

satisfying 
changing the trace of the Polyakov line. Seemingly, this 
overemphasizes the constraints, and one might fear that the 
algorithm is not ergodic. 

For this reason, we secondly constructed an algorithm where 
all links, ``participating'' in one particular Polyakov line, 
are updated simultaneously. We found that both algorithms 
yield the same results (at least for the observables shown below). 
Hence, ergodicity is not a problem for the present algorithms.

\section{ Numerical results } 
\begin{figure}
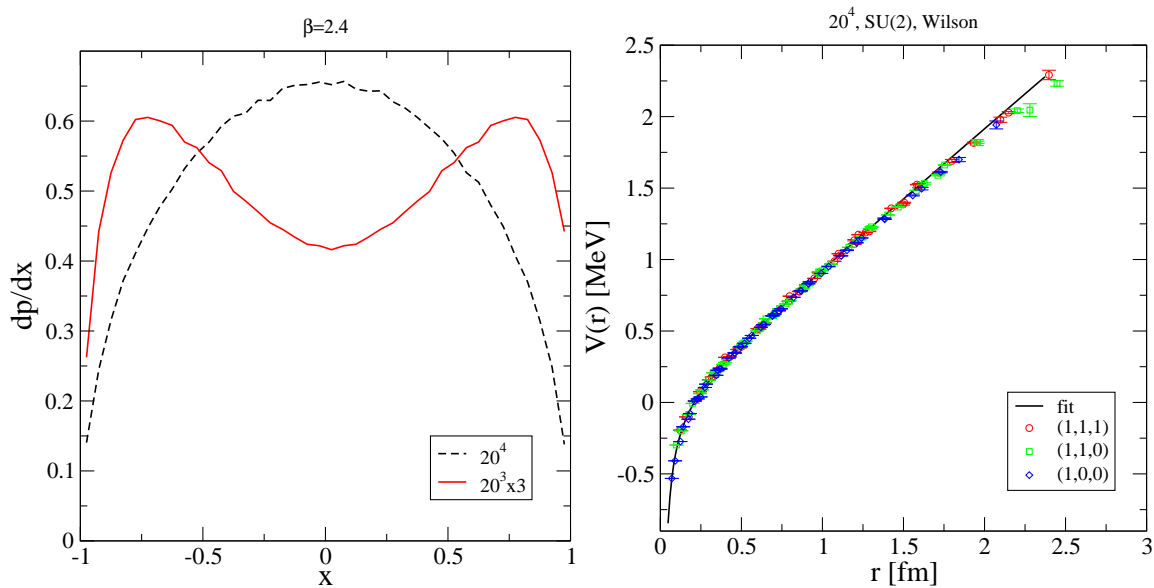

\centerline{
\epsfig{ file = polT.eps, 
width = 0.5 \textwidth } 
\epsfig{ file = pot_su2_nopol.eps, 
width = 0.5 \textwidth } 
}
\caption{ SU(2): Distribution of the Polyakov line expectation values  
for temperatures below and above the deconfinement one 
(left panel). Static quark anti-quark potential as function of 
the quark anti-quark distance for {\bf constrained } YM-theory 
at zero temperature (right panel). 
}
\label{fig:2} 
\end{figure}
In a first step, we have calculated the static quark anti-quark potential 
at zero temperatures by means of (spatially) smeared Wilson loops. 
Our findings for the constrained case are summarized in 
figure~\ref{fig:2}. The striking result is that, despite the constraints, 
the potential is still linear rising at large distances. In addition, 
the potential agrees for each $\beta $ separately with the potential 
from standard simulations. This implies that the scaling, i.e., 
the dependence of the lattice spacing with $\beta $, is unchanged by 
the constraints. In addition, the potential was obtained for 
different orientations of the quark anti-quark pair 
relative to the cubic lattice ((110), (110), (110) directions). 
Our result shows that breaking of rotational symmetry by the 
underlying lattice is small. 

\vskip 0.3cm 

\begin{figure}
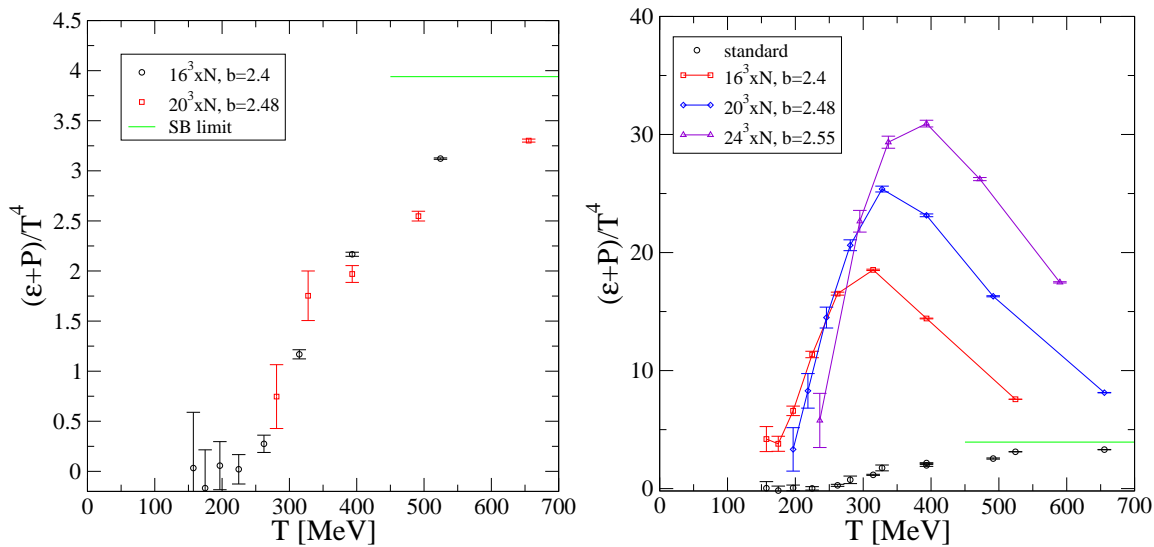

\centerline{
\epsfig{ file = energ_stan.eps, 
width = 0.5 \textwidth } 
\epsfig{ file = energ_all.eps,
width = 0.5 \textwidth } 
}
\caption{ SU(2): the sum of energy density $\epsilon $ and pressure $p$ 
as function of temperature. 
}
\label{fig:3} 
\end{figure}
Subsequently the question arises whether constrained YM-Mills theory 
possesses a deconfinement phase transition at high temperatures 
despite the constraints which set the Polyakov line expectation 
value to zero. In order to answer this question, we have monitored the 
``black body radiation'' of gluons as function of temperature. 
In particular, we have calculated 
$$
\frac{ \epsilon + p }{T^4 } \; = \; f(\beta) \; N_t^4 \;
( P_s - P_t)
$$
where  $\epsilon $ is the energy density and $p$ is the pressure, 
$P_s$, $P_t$ are space-like and time-like plaquettes, respectively, 
and $f(\beta )$ is a scaling function~\cite{Engels:1994xj}. 
Simulations were performed at fixed $\beta $ while temperature 
was varied by changing $N_t$. Our result for standard SU(2) YM-theory 
is shown in figure~\ref{fig:3} and agrees with the well-known 
results of the Bielefeld group~\cite{Engels:1994xj}: The onset 
of deconfinement is signaled by a sharp rise of $ \epsilon + p$. 
Figure~\ref{fig:3} also shows our result for {\it constrained } 
YM-Theory. We find also sharp rise of $ \epsilon + p$ at 
roughly the correct deconfinement temperature $T_c$. In addition, 
$ \epsilon + p$ strongly peaks at $T_c$ and seem to approach the 
Stefan Boltzmann limit from above. Unfortunately, we observe a 
strong cutoff dependence of $ \epsilon + p$  above $T_c$, which 
renders a physical interpretation of the results cumbersome. 

\vskip 0.3cm 
\begin{figure}
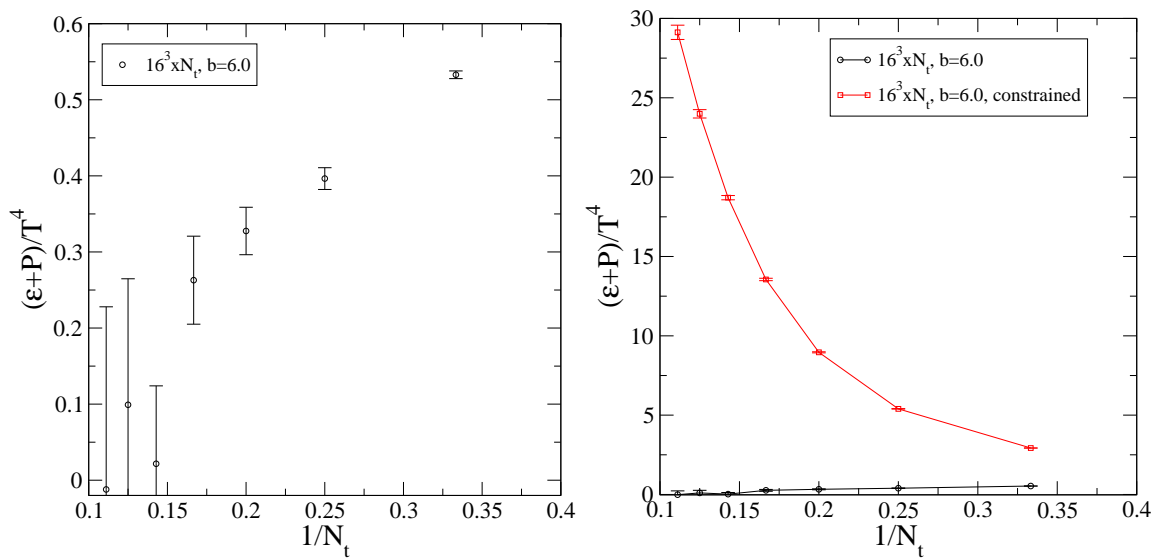

\centerline{
\epsfig{ file = energ_stan_su3.eps, 
width = 0.5 \textwidth } 
\epsfig{ file = energ_all_su3.eps, 
width = 0.5 \textwidth } 
}
\caption{ Same as figure 3 for the case of SU(3): 
standard SU(3) (left), constrained SU(3) (right). 
}
\label{fig:4} 
\end{figure}
Let us finally study the case of a SU(3) gauge theory with 
constraint (\ref{eq:csu3}). In figure~\ref{fig:4}, the  
result for $ \epsilon + p$ of standard YM-theory (left panel) 
is compared with that obtained from simulations of the constrained 
theory. In the constrained case we find that $ \epsilon + p$  
is monotonically decreasing with temperature. No signal 
of a deconfinement transition is observed. Moreover, it seems 
that black body radiation from deconfined gluons is present at all 
temperatures. For an interpretation of these findings, it is 
important to notice that SU(3) center symmetry is broken 
by the constraints. Also center symmetry is merely associated with 
quark confinement, there seems also no confinement of gluons 
if center symmetry is broken. 

\vskip 0.3cm
Acknowledgment: This work was supported in part by DFG-Re 856/5-1.

\end{document}